\documentclass{article}
\usepackage[dvips]{graphicx}

\makeatletter

\newcommand{\singlespacing}{\let\CS=\@currsize\renewcommand{\baselinestretch}{1}\tiny\CS}
\newcommand{\doublespacing}{\let\CS=\@currsize\renewcommand{\baselinestretch}{1.5}\tiny\CS}

\def\@citex[#1]#2{\if@filesw\immediate\write\@auxout{\string\citation{#2}}\fi
  \def\@citea{}\@cite{\@for\@citeb:=#2\do
    {\@citea\def\@citea{,\linebreak[0]\hskip0pt plus .2em}%
      \@ifundefined{b@\@citeb}%
        {{\bf ?}\@warning{Citation `\@citeb' on page \thepage\space undefined}}%
      \hbox{\csname b@\@citeb\endcsname}}}{#1}}

\makeatother

\begin{document}

\title{Evidence of crossover phenomena in wind speed data}

\author{Rajesh G. Kavasseri \\ Department of Electrical and Computer
Engineering \\ North Dakota State University, Fargo, ND 58105 -
5285 \\ ~(email: rajesh.kavasseri@ndsu.nodak.edu) \\   \\
Radhakrishnan Nagarajan \\  University of Arkansas for Medical
Sciences, Little Rock, AR 72205
 }

\date{}
\maketitle

\begin{abstract}
\noindent In this report, a systematic analysis of hourly wind
speed data obtained from three potential wind generation sites (in
North Dakota) is analyzed. The power spectra of the data exhibited
a power-law decay characteristic of $1/f^{\alpha}$ processes with
possible long-range correlations. Conventional analysis using
Hurst exponent estimators proved to be inconclusive. Subsequent
analysis using detrended fluctuation analysis (DFA) revealed a
crossover in the scaling exponent ($\alpha$). At short time
scales, a scaling exponent of $\alpha \sim 1.4$ indicated that the
data resembled Brownian noise, whereas for larger time scales the
data exhibited long range correlations ($\alpha \sim 0.7$). The
scaling exponents obtained were similar across the three
locations. Our findings suggest the possibility of multiple
scaling exponents characteristic of
multifractal signals.  \\
\end{abstract}


\section{Introduction}
Wind energy is a ubiquitous resource and is a promising
alternative to meet the increased demand for energy in recent
years. Unlike traditional power plants, wind generated power is
subject to fluctuations due to the intermittent nature of wind.
The irregular waxing and waning of wind can lead to significant
mechanical stress on the gear boxes and result in substantial
voltage swings at the terminals, \cite{spectrumaug2003}.
Therefore, it is important to build suitable mathematical
techniques  to understand the temporal behavior and dynamics of
wind speed for purposes of modeling, prediction, simulation and
design. Attempts to identify the features of wind speed time
series data were described in \cite{haslett79} and
\cite{raftery82}. To our knowledge, the first paper to bring out
an important feature of wind speed time series was
\cite{raferty_memory}. In \cite{raferty_memory}, the authors
examined long term records of hourly wind speeds in Ireland and
pointed out that they exhibited what is known as {\em long memory
dependence}. Seasonal effects, spatial correlations and temporal
dependencies were incorporated to build suitable estimators.
Trends in long term wind speed records were also suggested in
\cite{palutikoff91}. However, in \cite{raferty_memory}, short
memory temporal correlations were suggested by an examination of
the autocorrelation function. Evidence for the presence of long
memory correlations was provided by inspecting the periodogram of
the residuals from a fitted and autoregressive model of
order nine i.e. AR(9), \cite{raferty_memory}. \\

\noindent In this paper, we provide a systematic method to
identify, more importantly quantify the index of long range
correlations in wind speed time series data. We make use of a
fairly robust and powerful technique called {\em Detrended
Fluctuation Analysis} (DFA) in our analysis, \cite{peng1994}. The
rest of this paper is organized as follows. In Sec.2, the
acquisition of wind speed data is described. In Sec.3, traditional
analysis of the hourly wind speed using power spectral techniques
and Hurst estimators is discussed along with some of their
limitations. Detrended fluctuation analysis (DFA) is used to
capture the crossover phenomena in wind speed. Finally, the
conclusions are summarized in Sec. 4.

\section{Data Acquisition}
\begin{figure}[htbp]
\begin{center}
\includegraphics[height=3.5in,width=6in]{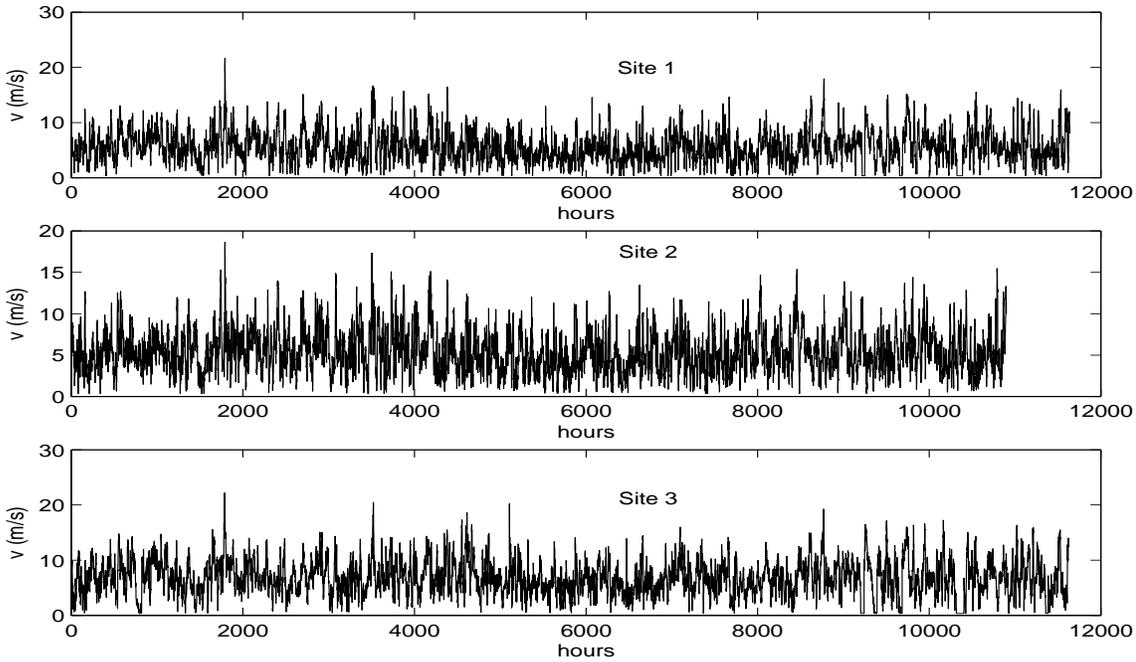}
\caption{Wind speed variations at the three locations.} \label{wspd}
\end{center}
\end{figure}

In this section, we provide a brief description of the wind speed
data acquisition system. The wind speeds at three different wind
monitoring stations in North Dakota are recorded by means of
conventional cup type anemometers located at a height of 20 m.
Wind speeds acquired every two seconds are averaged over a 10
minute interval to compute the 10 minute average wind speed. The
10 minute average wind speeds are further averaged over a period
of one hour to obtain the hourly average wind speed. In this
procedure, the computed hourly average wind speed is simply
equivalent to averaging the observations every two seconds for one
hour. The hourly average wind speeds are preferred over the 10
minute speeds to minimize storage requirements for several years
of data. The site details of the monitoring stations are provided
in Table 1. In our analysis, we consider a period from 11/29/2001
to 03/28/2003 for the wind speed records. Fig.\ref{wspd} shows the
wind speed ($v$ in $m/s$) variability at the three locations. The
aim of the present study is to characterize and quantify the
apparently irregular fluctuations of the wind speed in
Fig.\ref{wspd}. In the following section, the analysis of wind
speed data using three different techniques is provided.

\begin{small}
\begin{table}[htbp]
\label{tab1}
\begin{center}
%\caption{Station Locations}
\begin{tabular}{|c|c|c|c|}
\hline \hline Station & Latitude & Longitude & Elevation (ft) \\
\hline Site 1 & N 47 27.84' & W 99 8.18'& 1570 \\ \hline Site 2 & N
46 13.03' & W 97 15.10' & 1070 \\ \hline Site 3 & N 48 52.75' & W
103 28.44' & 2270 \\ \hline
\end{tabular}
\end{center}
\caption{ Station Locations}
\end{table}
\end{small}

\section{Analysis of wind speed data}

\subsection{Spectral Analysis}
\begin{figure}[!hptb]
\begin{center}
\includegraphics[height=3.5in,width=6in]{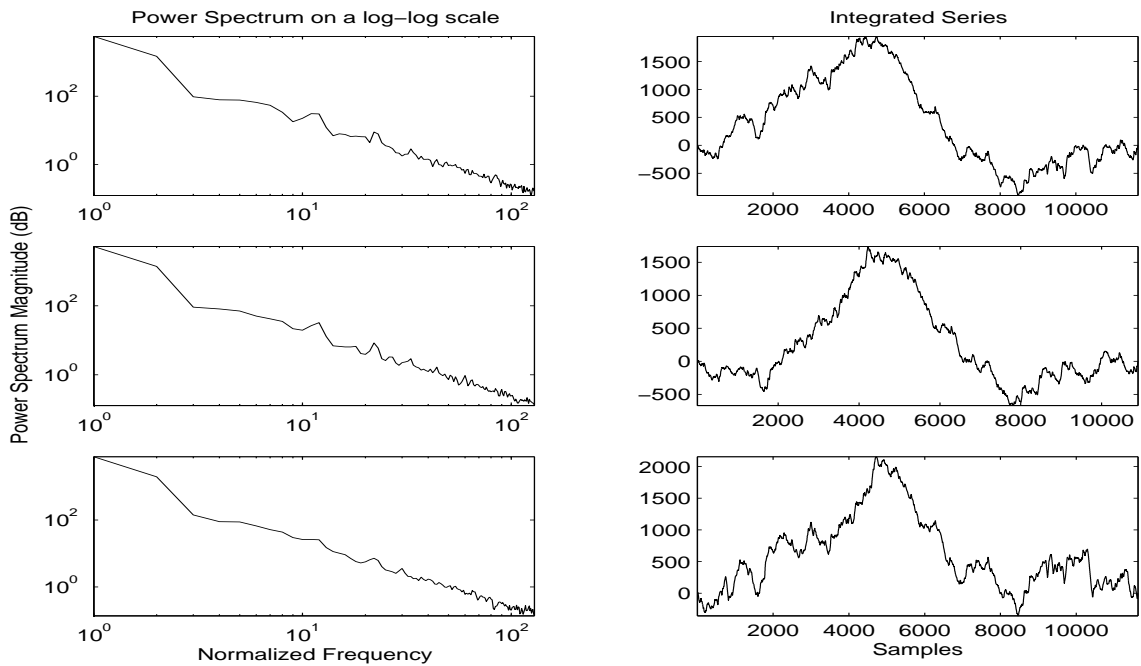}
\caption{On the left is the power spectrum  of the wind data
obtained at the three locations, Site 1 (top) , Site 2(middle) and
Site 3 (bottom) respectively. The corresponding integrated series is
shown on the right.} \label{prelim_figs}
\end{center}
\end{figure}

The power spectrum $S(f)$ shown in Fig. \ref{prelim_figs} exhibits
a power-law decay of the form $S(f) \sim 1/f^{\beta}$. The
auto-correlation functions (ACF) decay slowly to zero and the
first zero crossing of the ACFs occur at lags of 61, 56 and 60
hours respectively for the three data sets. Such features are
characteristic of statistically self-similar processes with well
defined long range power-law correlations, \cite{feder88}. In a
broad sense, long range correlations indicate that samples of the
time series that are very distant in time are correlated with each
other and can be captured by the auto correlation function or
equivalently, the power spectrum (as in Fig. \ref{prelim_figs}) in
the frequency domain. More precisely, a time series is self
similar if

\begin{equation}
y(t) \equiv a^{\alpha} y(t/a) \label{defn_selfsimilar}
\end{equation}

\noindent where $\equiv$ in Eqn. (\ref{defn_selfsimilar}) is used
to denote that both sides of the equation have identical
statistical properties. The exponent $\alpha$ in Eqn.
(\ref{defn_selfsimilar}) is called the self-similarity parameter,
or the scaling exponent. We note that while classical tools such
as autocorrelation functions, and spectral analysis can give
preliminary indications for the presence of long range
correlations, it may be difficult to use them unambiguously to
determine the scaling exponent. Additionally, these methods are
susceptible to non-stationary effects such as trends in the data
which are commonly encountered. It is thus important to seek
alternative measures that are potentially better suited than
classical tools to capture signal variability under different
temporal scales. In the following section, a few certain standard
methods in estimating long range correlations along with their
limitations is discussed.

\subsection{Hurst Exponents} \label{Hurst}
Hurst exponents have been successfully used to quantify long range
correlations in plasma turbulence \cite{plasmaturb1},
\cite{plasmaturb2}, finance \cite{Hurstfinance1}
\cite{Hurstfinance2}, network traffic, \cite{selfsimilar_traffic}
and physiology, \cite{ivanov_nature}. Long range correlations are
said to exist if $1/2 < H < 1$ (see \cite{beran}, \cite{bassing}
for details). There are several methods for the estimation of the
Hurst exponent such as variance method, re-scaled range (R/S),
analysis, periodogram method, Whittle estimator and wavelet based
methods. Table 1 shows the Hurst exponents estimated by various
methods using SELFIS, \cite{selfis} a tool for the analysis of
self-similar data  for all the three data sets. For the Whittle
and Abry-Veicht methods, $`int'$ refers to the 95 \% confidence
intervals in Table 2. In the case of variance method, re-scaled
range (R/S) and Absolute moments, $c$ refers to the correlation
coefficient in \%.

\subsubsection{Limitations of traditional Hurst estimators}
We note from Table 1 that, for a given method, the Hurst exponent
estimation is consistent across the three different data sets.
However, for a given data set, we note that there is no
consistency in the estimation of the exponents across the
different methods. While the Variance, R/S and Absolute moments
methods yield Hurst exponent estimates in the range (0.6-0.8), the
Whittle and Abry-Veicht methods yield exponents close to 1 and 1.4
respectively. The discrepancy in the results indicate potential
difficulties on applying Hurst estimators to experimental data
obtained from physical processes. Some of the estimators
implicitly assume stationarity of the data and can hence they
could be susceptible to nonstationarities, \cite{abryveitch98}.
Each of the Hurst estimators are derived under certain assumptions
(see \cite{taqqu}, \cite{beran} for details), and while they can
yield consistent estimates for synthetic data, there could be
discrepancies in the case of experimental data with trends. For
example, the R/S method, by the very nature of its construction
cannot be used to detect an exponent greater than one that may be
present in the data, \cite{kantelriver}. The Whittle estimator (a
parametric method based on the maximum likelihood estimator)
implicitly assumes that the parametric form of the spectral
density, or equivalently, the auto-correlation function is known.
An inappropriate choice of the parametric form could result in
potentially biased estimations, \cite{auditbacry}. Some of the
subtleties involved in traditional estimators can be also found in
\cite{plasmaturb2}. An additional problem arises if the given
process contains multiple scaling exponents under different
scaling regions in which case, Hurst exponent estimators are
difficult to apply directly. When multiple scaling exponents are
present, linear regression cannot be used to compute the exponent
over all scales and such an attempt can have adverse effects on
the Hurst exponent estimate as the regression may capture the
exponent over a certain scale. To summarize, the conflicting
estimates produced by these estimators indicates that one cannot
in general, use a ``blind" black-box approach when dealing with
processes with long range correlations. \\

\noindent It has also been pointed out that long-range
correlations can manifest themselves as slow moving non-stationary
trends, such as seasonal cycles, \cite{peng1995}. Thus traditional
techniques such as spectral analysis and Hurst estimators have
their limitations. These techniques are also not well suited to
provide insight in to possible change in the scaling indices
(crossover, Sec 3.3). Thus it is important to explore the choice
of alternate measures to quantify the scaling exponent from the
given data.
 In the following section, we shall present an over view
of such a method, i.e. Detrended Fluctuation Analysis (DFA).

\subsection{Detrended Fluctuation Analysis (DFA)}
The DFA first proposed in \cite{peng1994} is a powerful technique
and has been successfully used to determine possible long-range
correlations data sets obtained from diverse settings
\cite{hausdorff1995}, \cite{ausloos97}, \cite{ivanova_1},
\cite{peng1995}. A brief description of DFA is included here for
completeness. A detailed explanation can be found in elsewhere,
\cite{peng1994}. Consider a time series $\{x_k\}, k = 1,\dots N$.
Then, the DFA algorithm consists of the following steps.
\begin{itemize} \item Step 1  The series $\{x_k\}$ is
integrated to form the integrated series $\{y_k\}$ given by
\begin{equation} y(k) = \sum_{i=1}^{i=k} [x(i) - \bar{x}] \;\;\; k
= 1, \dots N
\end{equation}
\item Step 2  The series $\{y_k\}$ is divided in to
$n_s$ non-overlapping boxes of equal length $s$, where $n_s =
int(N/s)$. To accommodate the fact that some of the data points
may be left out, the procedure is repeated from the other end of
the data set and $2 n_s$ boxes are obtained, \cite{peng1994}.

\item Step 3  In each of the  boxes, the local trend is calculated
by a least-square fit of the series and the variance $F^2(v,s)$ is
calculated from

\begin{equation}
F^2(v,s) = \{ \frac{1}{s} \sum_{i=1}^{i=s} \{ y [(v-1)s + i] -
 y_v(i) \}^2
\end{equation}

 for each box $v = 1,\dots n_s$. Similarly, the computation is
 done for each box $v = n_s+1, \dots 2 n_s$ by
 \begin{equation}
 F^2(v,s) =
\{ \frac{1}{s} \sum_{i=1}^{i=s} \{ y [N-(v-n_s)s + i] -
 y_v(i) \}^2
 \end{equation}

 where $y_v$ is the fitting polynomial in box $v$. Depending on
 the polynomial, i.e. linear, quadratic, cubic, quartic, the
 procedure is called DFA1, DFA2, DFA3 and DFA4 respectively. The
 second order fluctuation is calculated by averaging the
 variations over each of the boxes , i.e.

\begin{equation}
 F_2(s) =
 \{ \frac{1}{2 n_s}  \sum_{v=1}^{v=2n_s}
 [F^2(v,s)] \}^{1/2}
 \end{equation}

\item Step 4 The computation in Step 3 is
repeated over various time scales by varying the box size $s$. A
log-log graph of the fluctuations $F_2(s)$ versus $s$ is
calculated . Linear relationships in the graph indicate
self-similarity and the slope of the line $F_2(s)$ vs $s$ on the
log-log plot determines the
scaling exponent $\alpha$. \\
\end{itemize}

\noindent The value of $\alpha$ obtained from the DFA algorithm
quantifies the nature of correlations. Values of $\alpha$ in the
range (0, 0.5) characterize anti-correlations (large fluctuations
are likely to be followed by small fluctuations and vice-versa)
and values of $\alpha$ in the range (0.5,1) characterize
persistent long range correlations (large/small fluctuations are
likely to be followed by large/small fluctuations in that order)
with $\alpha = 0.5$ representing uncorrelated (white) noise. If
$\alpha > 1$, correlations exist, but they are no longer of a
power law form, \cite{peng1995}. For exactly self-similar
processes, the exponent ($\beta)$ from the power spectrum ($S(f)
\sim 1/f^{\beta})$ is related to the DFA exponent $\alpha$ by
$\beta = 2\alpha -1$, \cite{beran}. For example, $\alpha = 0.5$ is
equivalent to $\beta = 0$ which characterizes white noise, while
$\alpha = 1$ is equivalent to $\beta = 1$ which corresponds to
$1/f$ noise and $\alpha = 1.5$ which corresponds to $\beta = 2$
characterizes Brown noise, the integration of white noise. However
in the case of experimental data which may be subject to trends
and non-stationarities, an unambiguous determination of scaling
exponents from the power spectrum may be difficult,
\cite{ekbunde}. DFA minimizes trends by local de-trending (Step 3)
and hence it is robust to trivial non-stationarities. While the
original DFA \cite{peng1994}, used only differencing of the
integrated series, Fig. 1, recent reports have pointed out that a
choice of higher order polynomial detrending can avoid spurious
results \cite{bunde2000}, \cite{hu2001}. Polynomial trends are
minimized by local detrending (Step 3). This renders DFA to be
robust to non-stationarities contributed by polynomial trends and
prevents spurious detection of long-range correlations which is an
outcome of such trends. The scaling exponents are estimated by
linear regression of the log-log fluctuation curve,
\cite{peng1994}. However, this can lead to spurious results when
there is more than one scaling exponent which is true in the case
of crossover phenomena, \cite{peng1994}. A crossover usually
arises due to changes in the correlation properties of the signal
at different temporal or spatial scales (see Figs.
\ref{dfa_alp01}, \ref{dfa_alp03} and \ref{dfa_alp05}),
\cite{hu2001}. Therefore, extracting the global exponent can be
misleading, especially in the presence of crossover phenomena
\cite{peng1995}, \cite{ausloos97}. Recent studies have suggested
comparing the results obtained on the original to constrained
randomized shuffles of the given data \cite{peng1995}. Unlike
traditional bootstrap realizations, constrained randomized
shuffles (surrogates) are obtained by resampling the given data
without replacement. In surrogate testing, one generates what are
termed as ``constrained realizations", \cite{theiler},
\cite{schreiber}. The constraint here is on retaining the
distribution of the original data in the surrogate realization.
While the temporal structure is destroyed, the distribution of the
original data is retained in the surrogate realization. The null
hypothesis addressed by the random shuffled surrogates (which
retain the pdf of the original data) is that the original data is
``uncorrelated".  The choice of the random shuffled surrogates
helps us to reject the claim that the observed scaling exponent is
due to the distribution as opposed to the correlation in the given
data. Comparison of the scaling exponents obtained on the original
data to that of the random shuffled surrogates is encouraged by
earlier reports \cite{hu2001}, \cite{goldberger},
\cite{hausdorff1995}, \cite{matia}, \cite{kantelriver}. A good
exposition on the concepts of surrogate analysis can be found in
\cite{theiler}, \cite{schreiber}, \cite{small}. \\

\noindent While the DFA has been shown to be a robust algorithm
compared to traditional Hurst exponent estimators in the presence
of non-stationarities, there are a few subtleties in the
application and interpretation of the results obtained from DFA. A
common problem with DFA is that crossovers can occur due to a
genuine change in the correlation properties of the signal, or due
to trends. While a choice of higher order polynomial de-trendings
can eliminate polynomial trends and avoid spurious results
\cite{bunde2000}, \cite{hu2001}, the presence of strong sinusoidal
trends can induce spurious crossovers, \cite{hu2001} (a detailed
discussion of this issue is reported in \cite{hu2001}). Since
trends are unavoidable in time series generated by physical
processes, it may be prudent to first recognize their presence
before applying DFA. For the data sets considered in this study we
did not observe strong sinusoidal trends and therefore pursued the
application of DFA, the results of which are described in the next
section.

\subsection{Results with DFA} 
\begin{figure}[htbp]
\begin{center}
\includegraphics[height=2in,width=4in]{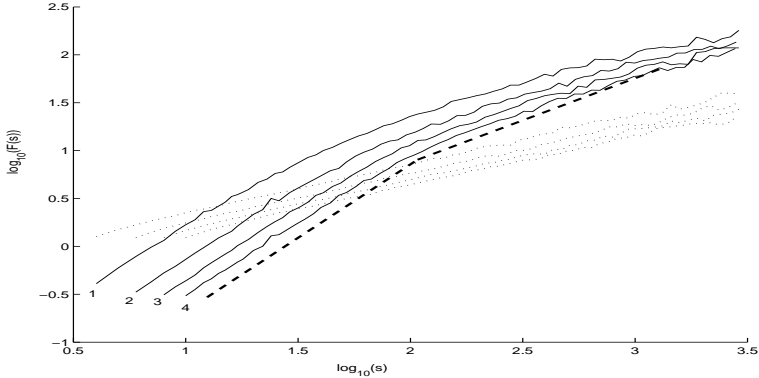}
%\caption{log-log plot of the fluctuation functions of the original
%data and its random shuffle surrogate (dotted lines) for Site 1.
%The order of polynomial detrending $d$ (1,2,3,4) for the original
%data set is indicated. The fourth order ($d=4$) detrending on the
%original data is shown by the dashed bold line.}
\caption{log-log plot of the fluctuation functions of the original
data and its random shuffled surrogate (dotted lines) for Site 1.
The order of polynomial detrending $d$ (1,2,3,4) for the original
data set is indicated. The fourth order ($d=4$) detrending on the
original data is shown by the dashed bold line.} \label{dfa_alp01}
\end{center}
\end{figure}

\begin{figure}[htbp]
\begin{center}
\includegraphics[height=2in,width=4in]{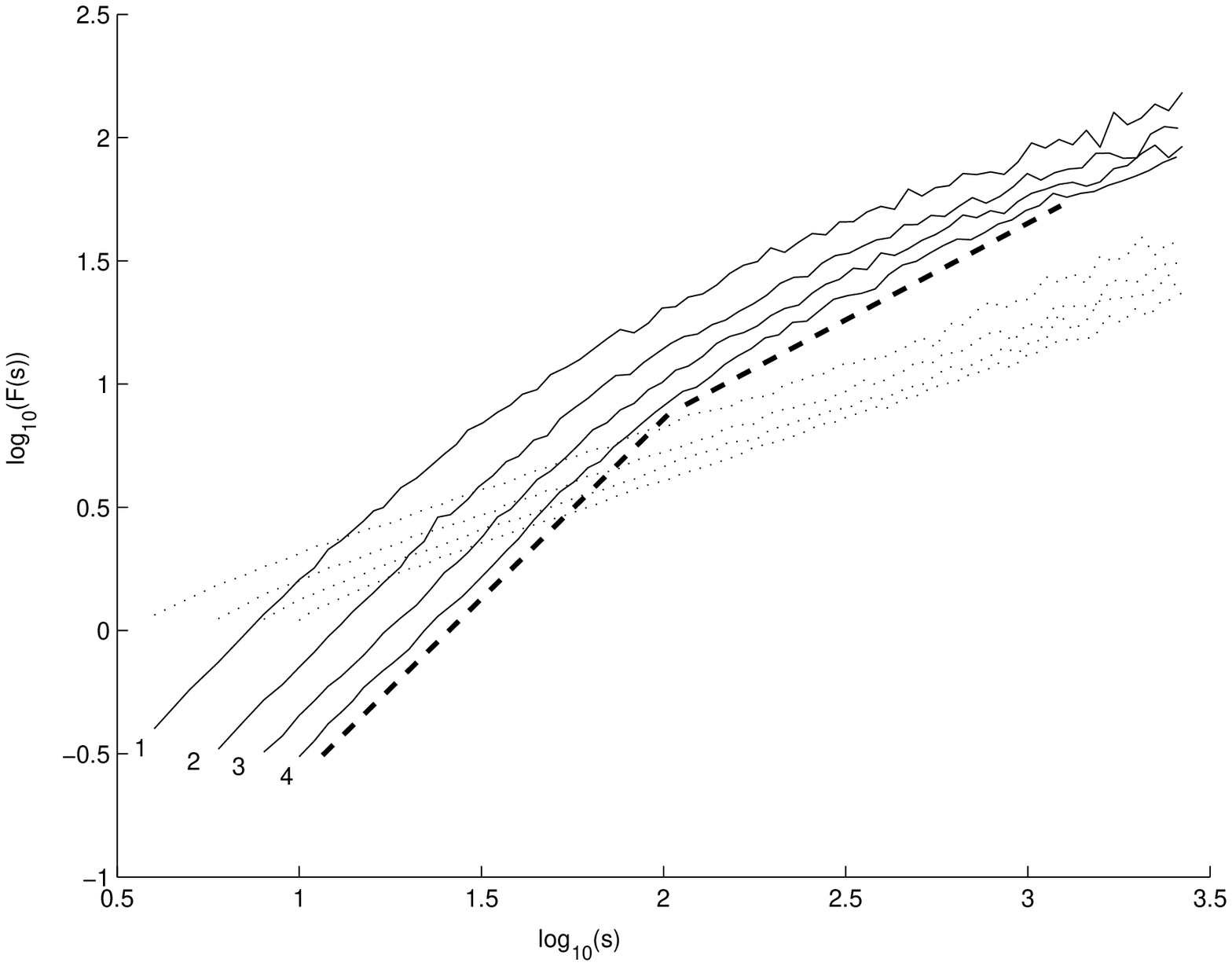}
\caption{log-log plot of the fluctuation functions of the original
data and its random shuffled surrogate (dotted lines) for Site 2.
The order of polynomial detrending $d$ (1,2,3,4) for the original
data set is indicated. The fourth order ($d=4$) detrending on the
original data is shown by the dashed bold line.} \label{dfa_alp03}
\end{center}
\end{figure}

\begin{figure}[htbp]
\begin{center}
\includegraphics[height=2in,width=4in]{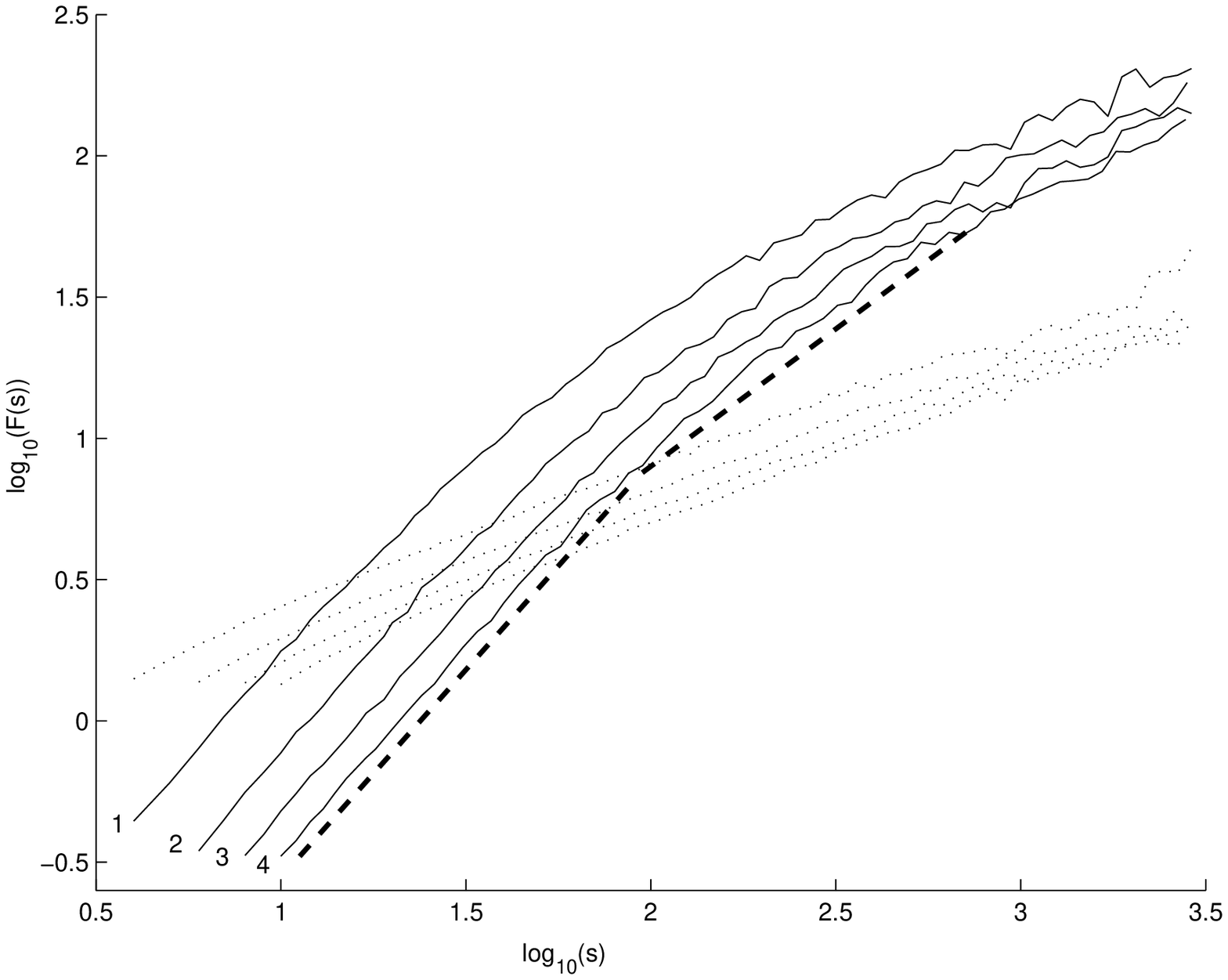}
\caption{log-log plot of the fluctuation functions of the original
data and its random shuffled surrogate (dotted lines) for Site 3.
The order of polynomial detrending $d$ (1,2,3,4) for the original
data set is indicated. The fourth order ($d=4$) detrending on the
original data is shown by the dashed bold line.} \label{dfa_alp05}
\end{center}
\end{figure}

The log-log plot of the fluctuation $F(s)$
versus the time scale $(s)$, for the three data sets and their
surrogates (indicated by the dotted lines) with different order
polynomial detrending (indexed by 1,2,3,4) are shown in Figs
\ref{dfa_alp01}, \ref{dfa_alp03} and \ref{dfa_alp05}. The scaling
exponents estimated by linear regression for all four orders of
detrending on the original data and its surrogates are summarized
in Table 3.

\begin{tiny}
\begin{table}[htbp]
\label{global_scale}
\begin{center}
%\caption{Hurst exponent estimation by various methods}
\begin{tabular}{|c|c|c|c|c|c|c|c|c|c|c|c|}
\hline \hline DATA & \multicolumn{2}{c|}{Variance} &
\multicolumn{2}{c|}{R/S} & \multicolumn{2}{c|}{Abs. Moments} &
\multicolumn{2}{c|}{Whittle} & \multicolumn{2}{c|}{Abry-Veicht} &
Period
\\ \cline{2-12}
 & $H$ & $c$ & $H$ & $c$ & $H$ & $c$ & $H$ & $int$ & $H$ & $int$ & $H$ \\ \hline
 Site 1 & 0.778 & 96.73 & 0.595 & 97.36 & 0.726 & 94.4 & 0.99 &
 0.98-1.01 & 1.36 & 1.34-1.38 & 1.028
 \\ \hline
 Site 2 & 0.772 & 97.3 & 0.608 & 97.35 & 0.712 & 95.5 & 0.99 &
 0.98-1.02 & 1.33 & 1.31-1.35 &  1.059
 \\ \hline
 Site 3 & 0.721 & 95.7 & 0.593 & 96.65 & 0.654 & 95.8 & 0.99 &
 0.98-1.02 & 1.36 & 1.34-1.37 &  1.077
 \\ \hline
\end{tabular}
\end{center}
\caption{Hurst exponent estimation by various methods}
\end{table}
\end{tiny}

\begin{table}[htbp]
\label{global_scale}
\begin{center}
%\caption{Global scaling Exponent of the Original data ($\alpha)$
%and it surrogates ($\alpha^*$)}
\begin{tabular}{|c|c|c|c|c|c|c|c|c|}
\hline \hline DATA & \multicolumn{2}{c|}{d=4} &
\multicolumn{2}{c|}{d=3} & \multicolumn{2}{c|}{d=2} &
\multicolumn{2}{c|}{d=1}
\\ \cline{2-9}
 & $\alpha$ & $\alpha^*$ & $\alpha$ & $\alpha^*$ & $\alpha$ & $\alpha^*$ & $\alpha$ & $\alpha^*$ \\ \hline
 Site 1 & 1.03 & 0.51 & 0.99 & 0.51 & 0.94 & 0.51 & 0.85 & 0.51
 \\ \hline
 Site 2 & 1.01 & 0.52 & 0.96 & 0.52 & 0.92 & 0.52 & 0.83 & 0.52
 \\ \hline
 Site 3 & 1.06 & 0.49 & 1.02 & 0.49 & 0.97 & 0.48 & 0.88 & 0.48
 \\ \hline
\end{tabular}
\end{center}
\caption{Global scaling exponent of the original data ($\alpha)$ and
it surrogates ($\alpha^*$)}
\end{table}

\begin{table}[htbp]
\label{local_scale}
\begin{center}
%\caption{Local Scaling Exponents of the Original data ($\alpha_1)$
%and  ($\alpha_2$)}
\begin{tabular}{|c|c|c|c|c|c|c|c|c|}
\hline \hline DATA & \multicolumn{2}{c|}{d=4} &
\multicolumn{2}{c|}{d=3} & \multicolumn{2}{c|}{d=2} &
\multicolumn{2}{c|}{d=1}
\\ \cline{2-9}
 & $\alpha_1$ & $\alpha_2$ & $\alpha_1$ & $\alpha_2$ & $\alpha_1$ & $\alpha_2$ & $\alpha_1$ & $\alpha_2$ \\ \hline
 Site 1 & 1.47 & 0.75 & 1.45 & 0.70 & 1.38 & 0.66 & 1.23 & 0.59
 \\ \hline
 Site 2 & 1.44 & 0.70 & 1.40 & 0.65 & 1.35 & 0.62 & 1.20 & 0.58
 \\ \hline
 Site 3 & 1.44 & 0.77 & 1.42 & 0.71 & 1.38 & 0.67 & 1.26 & 0.60
 \\ \hline
\end{tabular}
\end{center}
\caption{Local scaling exponents of the original data ($\alpha_1)$
and  ($\alpha_2$)}
\end{table}

\noindent From Table 3, we note that the choice of linear
detrending (DFA1, i.e. $d=1$) yields estimates of $\alpha \sim
0.85$ consistently at all three locations. Whereas, higher order
detrendings ($d=2,3,4$) indicate an exponent $\alpha \sim 1$
consistently at all three locations which suggests a possible
$1/f$ type behavior. For the surrogate data sets ($a^{*}$) in
Table 3, we note that all four choices of detrending yield
exponents very close to $0.5$ at all three locations which shows
that scaling in the original data is an outcome of the
correlations present in it and not due to its
distribution. \\

\noindent We further note from Figs. \ref{dfa_alp01},
\ref{dfa_alp03} and \ref{dfa_alp05} that unlike the surrogates,
the log-log plot of the original data sets at all three locations
is not linear. For low orders of detrending $d = 1,2,3$, the slope
of the fluctuation functions of the original data set gradually
changes as seen in Figs \ref{dfa_alp01}, \ref{dfa_alp03} and
\ref{dfa_alp05}. However, for $d =4$ (fourth order detrending),
the transition of slope is comparatively abrupt in the fluctuation
function around $s_{\times} \sim 10^2$ which suggests the
existence of more than one scaling exponent. In this case, the
global scaling exponent shown in Table 3 is insufficient to
capture the change in scaling exponent. Therefore, the scaling
region is partitioned in to two regions around $s = s_{\times}$.
In the region $s < s_{\times}$, the slope of the fluctuation
function is given by $\alpha_1$ and in the region $s >
s_{\times}$, the slope is given by $\alpha_2$. These represent
what we call ``local scaling exponents". Thus, the ``crossover"
from one scaling exponent ($\alpha_1$) to the other ($\alpha_2$)
is seen to occur at approximately at a time scale $s_{\times} \sim
10^2 = 100$ hours. The local scaling exponents $( \alpha_1,
\alpha_2)$ estimated by DFA for these two regions, using different
order polynomial detrending for the three data sets are summarized
in Table 4. As a comment, we would like to mention that such
intricacies are not evident from the power spectrum, Fig.
\ref{prelim_figs} and Hurst
analysis (Sec 3.2).\\

\noindent Recall from Sec. \ref{Hurst} that while some of the
methods (Variance method, R/S and Absolute moments) yielded
exponent estimates in the range (0.6-0.8), the Abry-Veicht method
produced an estimate close to 1.4. On the other hand, DFA in
addition to identifying the exponents also helps in demarcating
the regions
(time scales in the signal) where these exponents are contained.\\

\noindent From Table 4, we note that DFA1 yields an exponent
$\alpha_1$ close to 1.2 whereas DFA2,3,4 yield exponents close to
1.45 at all three locations. For the exponent $\alpha_2$, we note
that while DFA1 yields estimates close to 0.6, DFA2,3,4 yield
exponents close to 0.7. Therefore, it is reasonable to say that
the original signal possesses {\em at least} two scaling exponents
over two time scales. At short time scales ($ s < 100 $ hours),
the data exhibits behavior similar to Brown noise $( \alpha \sim
1.45)$, whereas for longer time scales $(s >> 100$ hours) one
observes persistent long-range correlations $( \alpha \sim 0.70)$.
Interestingly, the results obtained across the data obtained from
different geographical locations seem to exhibit a similar
behavior.

\section{Conclusions}
Long term records of hourly average wind speeds at three different
wind monitoring stations in North Dakota are examined. Preliminary
spectral analysis of the data indicates that wind speed time
series contain long range power-law correlations. Analysis using
Hurst estimators were inconclusive. A detailed examination using
{\em DFA} indicated a crossover and revealed the existence of {\em
at least} two distinct scaling exponents over two time scales.
While the data resembled Brownian noise over short time scales,
persistent long range correlations were identified over longer
time scales. The scaling behavior was consistent across the three
locations and were verified using different orders of polynomial
trending. It is interesting to note that despite the inherent
heterogeneity across spatially separated locations, certain
quantitative features of the wind speed are retained. While
several factors including friction, topography and surface heating
are known to contribute towards wind speed variability, the
present report seems to indicate that the combined effect of these
factors may themselves be subject to variability over different
time scales. A possible explanation for the crossover may be that
on short time scales (tens of hours), the fluctuations in wind
speed may be dominated by atmospheric phenomena governed by the
``local or regional" weather system whereas on longer time scales
(extending from several days to months), the fluctuations may be
influenced by more general ``global" weather patterns. While the
present study indicates two distinct scaling exponents, a closer
inspection of longer records of wind speed at finer resolutions
may possibly reveal a spectrum of scaling exponents characteristic
of multifractals. However, this is more of a conjecture at this
point. We plan to investigate this in greater detail in subsequent
studies.

\section*{Acknowledgment}
\noindent We thank the reviewers for their constructive comments
and useful suggestions which have helped us enhance the quality of
the manuscript. The financial support from ND EPSCOR through NSF
grant EPS 0132289 and services of the North Dakota Department of
Commerce : Division of Community Services are gratefully
acknowledged.

\end{document}